# Hierarchy of Domain Reconstruction Processes due to Charged Defect Migration in Acceptor Doped Ferroelectrics

Ivan S. Vorotiahin,[1,†] Anna N. Morozovska,[2] and Yuri A. Genenko[1,*]

[1]Institute of Materials Science, Technische Universität Darmstadt, Otto-Berndt-Str.3, Darmstadt, Germany

[2]Institute of Physics, National Academy of Sciences of Ukraine, pr. Nauky 46, Kyiv, Ukraine

Evolution of a stripe array of polarization domains triggered by the oxygen vacancy migration in an acceptor doped ferroelectric is investigated in a self-consistent manner. A comprehensive model based on the Landau-Ginzburg-Devonshire approach includes semiconductor features due to the presence of electrons and holes, and effects of electrostriction and flexoelectricity especially significant near the free surface and domain walls. A domain array spontaneously formed in the absence of an external field is shown to undergo a reconstruction in the course of the gradual oxygen vacancy migration driven by the depolarization fields. The charge defect accumulation near the free ferroelectric surface causes a series of phenomena: (i) symmetry breaking between the positive and negative c-domains, (ii) appearance of an effective dipole layer at the free surface followed by the formation of a surface electrostatic potential, (iii) tilting and recharging of the domain walls, especially pronounced at higher acceptor concentrations. An internal bias field determined by the gain in the free energy of the structure exhibits dependences of its amplitude on time and dopant concentration well comparable with available experimental results on aging in $BaTiO_3$.

[†]Corresponding Author: isvort@mm.tu-darmstadt.de
[*]Corresponding Author: genenko@mm.tu-darmstadt.de

# 1   Introduction

One of the long-standing problems related to reliability of ferroelectrics is the ubiquitous phenomenon of aging well known since the fifties [1, 2, 3], which is mostly pronounced in acceptor doped ferroelectrics. Aging is the gradual change of material properties at constant thermodynamic conditions on a typical time scale of $10^2$-$10^7$ s. Typically, aging in ferroelectrics reveals itself through the shift or deformation of the polarization-field loop, as well as decrease of maximum polarization, strain and permittivity. Mason suggested in 1955 a relaxation mechanism of aging by slow thermally activated motion of domain walls [2], while Plessner explained a logarithmic decay of properties by a wide distribution of activation energies, reflecting the random character of obstacles to domain wall motion in a ceramic [3]. Further studies were concentrated on microscopic mechanisms behind the aging. Initially a popular hypothesis of materials degradation was an ionic space charge accumulation at charged domain boundaries [4, 5, 6]. Alternatively, it was proposed that a gradual alignment of polar defects stabilizes domain structures [7, 8, 9]. The latter two mechanisms become most common qualitative explanations of a variety of aging phenomena, comprehensively presented in reviews [10, 11], however, models quantifying these processes were lacking long time.

The first quantitative model was advanced by Arlt and Neumann in 1988 who showed that, for energy reasons, polar defects, consisting of an acceptor impurity on the B-site of a perovskite $ABO_3$ and an adjacent oxygen vacancy, tend to align with the polarization direction of the surrounding lattice by thermally activated vacancy jumps over the oxygen octahedron [ 12 ]. The gradual statistical orientation of such defect dipoles in a macroscopically unpoled state entails freezing of this state and prevents its switching to the macroscopically polarized state that is manifested by pinching of the polarization-field hysteresis loop. For the initially macroscopically poled system the polarization loop is gradually shifted along the field axis. In both, poled and unpoled, cases aging can be quantified by a so called "internal bias field" $E_{ib}$ [9]. The theory was able, in principal, capture experimental time and temperature dependences of $E_{ib}$ observed in Ni-doped $BaTiO_3$ [12]. A weakness of the theory consisted in a very rough electrostatic estimation of the energy barrier for the oxygen jumping between the sites on the oxygen octahedron, which plays a crucial role for the dipole orientation kinetics. Concerning this matter, a significant progress has been recently achieved in the microscopic understanding of the defect dipole formation and orientation by density functional theory (DFT) calculations of the defect energies and relevant barriers [13, 14, 15]. The latter approach was further combined with advanced microscopic methods for identification of the presence, orientation and position of dipoles with respect to domain walls, *e.g.* by electronic paramagnetic resonance (EPR) [16, 17, 18]. These studies revealed different values of the energy barriers depending on phase symmetries and different dopant ions in the range from 0.24 to 1.4 eV. Based on the microscopically evaluated energy values the kinetics of dipole orientation was studied by solving appropriate rate equations in various field and polarization regimes [14].

Previous [9, 6, 12] and more recent [19, 20, 21, 22, 23] studies revealed the range of activation energies relevant to aging from 0.4 eV to 1.1 eV. Shorter characteristic times and smaller energies about 0.4-0.7 eV were extracted from de-aging experiments restoring the virgin (unaged) state by application of an AC field [9, 24] or by heating above the Curie temperature to a paraelectric state [21]. Analysis of experimental data from original experiments and literature performed by Morozov and Damjanovic [22, 23] revealed that the shorter characteristic times and smaller energies involved in de-aging and accompanied by opening of the contracted "aged" polarization loops are rather related to the short-range rearrangements of the oxygen vacancies like the "cage motion" within one oxygen octahedron. The related energies and aging times are in agreement with DFT calculations [13 – 15] and kinetic simulations [14]. On the other hand, the further opening of the pinched polarization loops up to their virgin, rectangular form [23] as well as the shifting of the polarization loop in the poled state require much longer times [25] and involve activation energies from 0.9 eV to 1.1 eV. Such energies are characteristic of ionic dc-conductivity [23] and are related to long-range migration of oxygen vacancies as was proven by Waser [26].

Important limitation of the defect-dipole orientation model is that it ignores interaction between dipoles and therefore cannot explain either the saturating dependence of the internal bias field or the monotonic decrease of the characteristic aging time with increasing doping concentration observed in experiments [9, 12, 27]. This can, however, be explained by the above-mentioned space-charge scenario being intrinsically a collective mechanism. This concept was for a long time quantitatively realized only in oversimplified one-dimensional (1D) models [6, 28]. More elaborate two-dimensional (2D) [29, 30] and three-dimensional (3D) [26] space-charge mechanism models allowed reasonable explanations of the time, temperature and concentration dependences of the internal bias field at long observation times, related to long-range defect migration, as well as the observed decrease of the characteristic aging time with increasing doping concentration [9, 12, 27].

Though the essential features of aging concerning its time, temperature and doping dependence are mainly captured by the actual defect-dipole and space-charge theories, both concepts remain unrealistic in one assumption: they assume that oxygen-vacancy short- or long-range migration occurs within a fixed domain pattern. In fact, experimental visualization clearly revealed the evolution of domain patterns with time [31, 32], and it is conceivable that domains undergo evolution together with the defect rearrangement. To account for these cooperative processes a very well elaborated model is required that includes defect and domain interactions on the mesoscopic scale of at least several domains. Such analysis can be performed by means of the time-dependent Landau-Ginzburg-Devonshire (T-LGD) approach also called phase-field modelling [33, 34, 35]. In this study, this approach is applied to the stripe domain structures arising in thin ferroelectric films.

A deep physical understanding and possible control of polar properties are important for both fundamental research and wide range of advanced applications of ferroelectric thin films in nonvolatile memories, nanoelectronics and sensorics [36, 37]. With decreasing the film thickness its ferroelectric properties usually decline until their complete disappearance at thicknesses smaller than the critical one [38]. Feasible ways to avoid the size-induced phase transition in thin epitaxial films are, for example, selecting of an appropriate substrate [39, 40] or modification of their chemical composition [41] including the incorporation of suitable defects.

Ferroelectric phase stability requires effective screening of the polarization bound charge at surfaces and interfaces [42]. As a rule, a theoretical analysis of bulk ferroelectric state assumes either complete screening of polarization by the electrodes, or considers the emergence of multidomain states as a pathway to minimize depolarization field energy [43]. Rapid growth of ferroelectric thin film applications urgently requires the analysis of nonlinear polarization dynamics stemming from non-zero spatial separation between bound and screening charges [44]. The spatial separation is often introduced via non-ferroelectric dead layers [45] and/or physical gaps [46, 47] separating the ferroelectric surface from the electrode. Because of the long-range nature of depolarization fields the incomplete surface screening of ferroelectric polarization leads to a nontrivial thermodynamics and kinetics of the domain structure [42, 43]. These in turn cause unusual phenomena near the electrically opened surfaces such as correlated polarization switching [48], formation of flux-closure domains in multiaxial ferroelectrics [49, 50, 51], domain wall broadening [50, 51, 52], the crossover between nonlinear screening regimes in ferroelectric films [53], labyrinthine domain patterns [54, 55] and meandering instabilities of anferrodistortive-ferroelectric domain walls in multiferroics [56, 57].

Notably, the stabilization of the ferroelectric state in thin ferroelectric films can take place due to chemical switching [58, 59, 60, 61], so that the screening via ionic adsorption is intrinsically coupled to the dynamics of surface electrochemical processes [62, 63]. The presence of point defects (such as charged or neutral; impurities and vacancies), acting as elastic dipoles, can strongly impact the electric polarization of the films via electrostriction and flexoelectric effect [64, 65, 66]. However, the domain structure and polar state of thin films in contact with atmosphere under the presence of elastic (neutral or charged) defects are very poorly studied, albeit the total volume of research in this area is extremely wide. Consequently, a comprehensive theoretical approach has not yet been developed.

In this work we consider aging processes in a self-organized stripe structure of polarization domains in an acceptor doped thin ferroelectric film using the time-dependent LGD approach. The details of the model including the free energy functional, basic equations and boundary conditions are disclosed in section 2 and a related **Appendix A**. The physical results of the numerical analysis and analytical estimations are presented and discussed in section 3. Main conclusions are summarized in section 4.

## 2 Problem Statement

Let us consider a thin ferroelectric film of thickness $h$, with electrically open top surface, and a bottom surface attached to a rigid electrode. The top electrode is separated from the film surface by the distance $d$. The film contains charged defects, namely, immobile acceptor impurities and mobile (donor) oxygen vacancies. The domain structure of the film arises spontaneously and rapidly, and then the migration of the mobile defects becomes the main mechanism of the further "self-organized" polarization evolution. The goal of this work is to establish the influence of the defect migration on the ferroelectric properties of the film.

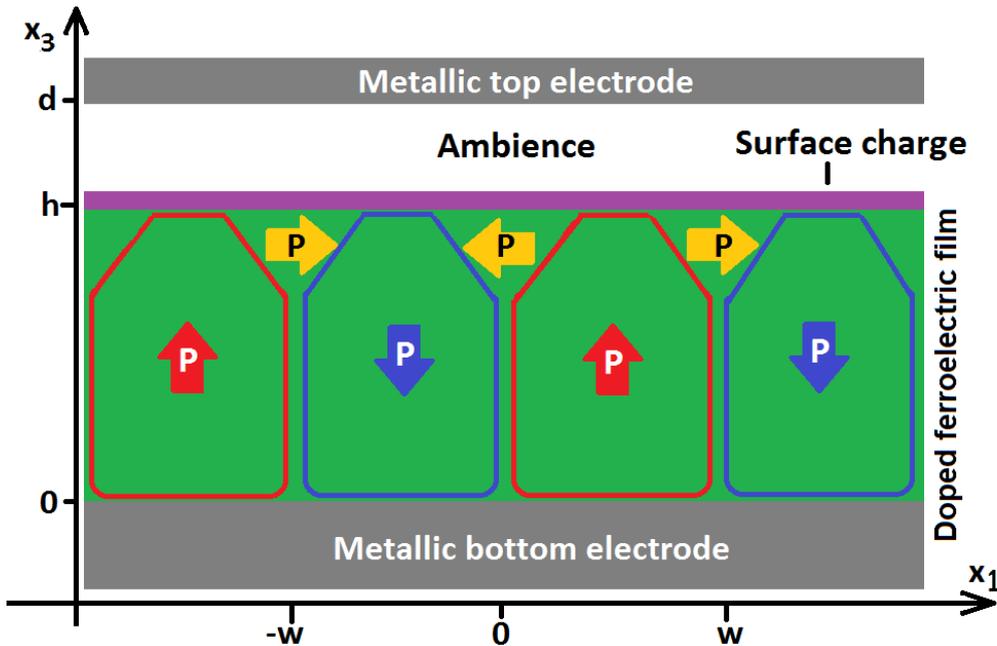

Fig. 1. Schematic image of a homogeneously doped ferroelectric film of thickness $h$, with domains of width $w$ with vertically oriented polarization (red and blue frames with arrows) and closure domains with laterally oriented polarization (yellow arrows). The film is clamped to a rigid conductive bottom electrode, has a layer of surface screening charge on the top surface and is separated by a thick ambient layer of thickness $d$ from the top electrode.

## 2.1 Free Energy of the System

The Gibbs thermodynamic potential for the ferroelectric film in relation to polarization **P**, electric potential $\varphi$ and electric field **E**, mechanical stress **σ**, electron $n$ and hole $p$ concentrations, donor $N_d^+$ and acceptor $N_a^-$ concentrations is considered as a sum of the bulk and surface contributions. The bulk part of the Gibbs free energy, for materials with an inversion center of symmetry in their parent phase, has the following form [64]:

$$G_V = \int_V d^3r \begin{pmatrix} \dfrac{\alpha_{ik}}{2} P_i P_k + \dfrac{\beta_{ijkl}}{4} P_i P_j P_k P_l + \dfrac{\gamma_{ijklmn}}{6} P_i P_j P_k P_l P_m P_n + \dfrac{g_{ijkl}}{2}\left(\dfrac{\partial P_i}{\partial x_j}\dfrac{\partial P_k}{\partial x_l}\right) - P_i E_i \\ -Q_{ijkl}\sigma_{ij} P_k P_l - \dfrac{s_{ijkl}}{2}\sigma_{ij}\sigma_{kl} - F_{ijkl}\sigma_{ij}\dfrac{\partial P_l}{\partial x_k} + e\varphi(Z_d N_d^+ - Z_a N_a^- - n + p) \\ -N_d^+ E_d - N_a^- E_a - T(S_d[N_d^+] + S_a[N_a^-]) + nE_C + pE_V - T(S_e[n] + S_h[p]) \\ + \dfrac{3k_B T}{2}\left(N_C F_{3/2}\left(\dfrac{E_g + e\varphi}{k_B T}\right) + N_V F_{3/2}\left(\dfrac{E_g - e\varphi}{k_B T}\right)\right) \end{pmatrix} \quad (1)$$

and contains several contributions, namely, Landau energy with a gradient term, electrostriction contribution, elastic energy, flexoelectric energy, electrostatic energy, donor, acceptor, electron and holes contributions along with their entropies.

The coefficients of the Landau expansion by the powers of polarization are $\alpha_{ik}$, $\beta_{ijkl}$ and $\gamma_{ijklmn}$, $g_{ijkl}$ is the gradient coefficient tensor. $\sigma_{ij}$ is the elastic stress tensor, $Q_{ijkl}$ is the electrostriction tensor, $F_{ijkl}$ is the flexoelectric effect tensor, $s_{ijkl}$ is the elastic compliance. $e$ is the elementary charge, $Z_d$ and $Z_a$ are the donor and acceptor ionization degrees, respectively. Only ionized donors (e.g. impurity ions or oxygen vacancies) are regarded mobile while acceptors are immobile. $E_d$ is the donor level, $E_a$ is the acceptor level, $E_C$ is the bottom of the conduction band, $E_V$ is the top of the valence band. The entropies of ionized donors and acceptors are denoted as $S_d[N_d^+]$ and $S_a[N_a^-]$, and the entropy of electron and hole Fermi gas, $S_e[n]$ and $S_h[p]$ [64]. The last term in Eq.(1), is the electron and hole kinetic energy, where $F_{3/2}(\xi)$ is the Fermi 3/2-integral with its general form given below. All quantities and coefficients included into Eq.(1) are listed in Table AI (see **Appendix A**) for the case of BaTiO$_3$ film.

Contributions into the free energy, originated from the entropies of defects are [64, 67, 68]:

$$S_Z = -k_B\left(N_Z^0\left(\left(\dfrac{N_Z^s}{N_Z^0}\right)\ln\left(\dfrac{N_Z^s}{N_Z^0}\right) + \left(1 - \dfrac{N_Z^s}{N_Z^0}\right)\cdot\ln\left(1 - \dfrac{N_Z^s}{N_Z^0}\right)\right)\right) \quad (2)$$

Here subscript $Z=d$ for donors, and $Z=a$ for acceptors, and superscript $s="+"$ for donors, and $s="-"$ for acceptors

$$S_e = -k_B N_C \int_0^{n/N_C} F_{1/2}^{-1}(\tilde{n}) d\tilde{n}, \qquad S_h = -k_B N_V \int_0^{p/N_V} F_{1/2}^{-1}(\tilde{p}) d\tilde{p} \qquad (3)$$

. Density of states in the conduction and valence bands can be found from electron and hole effective masses using the conventional expression, $N_{C,V} = \left(\dfrac{m_{n,p} k_B T}{2\pi\hbar^2}\right)^{3/2}$. $F_{1/2}^{-1}(\psi)$ is the inverse Fermi integral for an index ½. The Fermi integral itself is generally introduced as follows: $F_j(\psi) = \dfrac{1}{\hat{\Gamma}(j+1)} \int_0^\infty \dfrac{\omega^j d\omega}{1 + \exp(\omega - \psi)}$, where $\hat{\Gamma}(j)$ is the Euler Gamma-function.

We assume natural boundary conditions corresponding to zero energy of film surfaces.

## 2.2 Basic equations

In order to account for all properties of interest in the thin film of barium titanate, we use a system of equations containing the time-dependent LGD (other name Landau-Khalatnikov) equation for polarization, the Poisson equation for electrostatic potential, the continuity equation for each type of charge carriers and defects, the generalized Hooke law for elastic stress, and the elastic equilibrium equation. This allows us to account for electromechanical and semiconductor properties of the film focusing attention at the charge redistribution within its bulk.

The variation of the Gibbs free energy functional by polarization gives three T-LGD equations:

$$\Gamma \frac{\partial P_i}{\partial t} = -\frac{\delta G_V}{\delta P_i} \qquad (4)$$

where $\Gamma$ is the Khalatnikov constant. Eqs.(4) are supplied by the natural boundary conditions on the top and the bottom surfaces respectively:

$$\left. \left( g_{ijkl} \frac{\partial P_i}{\partial x_j} - F_{ijkl} \sigma_{ij} \right) \right|_{x_3 = 0, h} = 0, \qquad (5)$$

The electrostatic properties of the material can be described using the Poisson equation for the electric potential $\varphi$ related to the electric field $E_i = -\nabla_i \varphi$. It contains terms correspondent to the contributions from the crystal lattice itself, ferroelectric polarization, and the space charge caused by electronic charge carriers of both signs, as well as donor and acceptor dopants. The distribution of $\varphi$ can be devised both for the film bulk and for the ambience, where the space charge and the ferroelectric contribution are absent. Thus, for the film the Poisson equation reads as

$$\varepsilon_0 \varepsilon_b \frac{\partial^2 \varphi}{\partial x_i \partial x_i} = \frac{\partial P_j}{\partial x_j} - e(Z_d N_d^+ - Z_a N_a^- + p - n) \qquad (6a)$$

while in the ambience it turns into the Laplace equation,

$$\varepsilon_0 \varepsilon_e \frac{\partial^2 \varphi}{\partial x_i \partial x_i} = 0 \qquad (6b)$$

with boundary conditions setting up the properties of the bottom electrode (ground), the interface between the film and the ambience, and the top electrode:

$$\varphi|_{x_3=0} = 0, \quad \left. \left( D_n^{ext} - D_n^{int} + \varepsilon_0 \frac{\varphi}{\lambda} \right) \right|_{x_3=h} = 0, \quad \varphi|_{x_3=h-0} = \varphi|_{x_3=h+0}, \quad \varphi|_{x_3=h+d} = 0 \qquad (7)$$

where the electric displacement is defined by $\mathbf{D}=\varepsilon_0\varepsilon_b\mathbf{E}+\mathbf{P}$, and $\lambda$ is an effective screening length [69, 70].

In order to accomplish the Poisson equation (6a), we should account for all charge carriers and defects present in the ferroelectric film. In this model, the film is assumed to be uniformly doped with immobile acceptor ions of a constant concentration $N_a^-$, which remain negatively double-charged throughout the overall charge redistribution process. To warrant the charge neutrality, acceptors are assumed to be compensated by the mobile donor ions (*e.g.* oxygen vacancies) which also remain positively double-charged throughout the charge migration process and are initially uniformly distributed with the concentration $N_d^+$ equal to $N_a^-$. Under these circumstances, the initial concentrations of electrons and holes in the film are negligibly small and remain further very small due to the assumed blocking boundary conditions at the bottom electrode. Nevertheless, they are consistently accounted for in the following consideration.

To describe the evolution of charge carrier and defect densities the continuity equations are used for each type of charged species. Blocking electrodes are considered, which means that neither electronic charge carriers nor defects can migrate across the film surfaces. Solutions of the continuity equations are concentrations of charged species, which contribute to the space charge density expression in the Poisson equation (6a).

The continuity equation for electrons is formulated using the drift-diffusion approximation:

$$\frac{\partial n}{\partial t} - \frac{1}{e}\frac{\partial J_i^e}{\partial x_i} = 0 \tag{8}$$

and contains an expression for the electron current, $J_i^e = e\eta_e n E_i + \eta_e k_B T \nabla_i n$, characterized by the value of electron mobility $\eta_e$. Boundary conditions at the blocking electrode and at the top free surface are $J_3^e\big|_{x_3=0} = 0$ and $J_3^e\big|_{x_3=h} = 0$.

The equation for holes is analogous to the one for electrons:

$$\frac{\partial p}{\partial t} + \frac{1}{e}\frac{\partial J_i^h}{\partial x_i} = 0. \tag{9}$$

Here the hole current in the drift-diffusion approximation, $J_i^h = e\eta_h p E_i - \eta_h k_B T \nabla_i p$, contains the value of hole mobility $\eta_h$. Boundary conditions at the blocking electrode and at the top free surface are $J_3^h\big|_{x_3=0} = 0$ and $J_3^h\big|_{x_3=h} = 0$.

Donor concentration is described by the continuity equation of the following form:

$$\frac{\partial N_d^+}{\partial t} + \frac{1}{eZ_d}\frac{\partial J_i^d}{\partial x_i} = 0. \tag{10}$$

The donor current that contains donor mobility $\eta_d$ and is restricted, due to steric limitations [66], by the maximum possible donor concentration $N_d^0$ has the following expression [67, 68]:

$$J_i^d = eZ_d \eta_d N_d^+ E_i - \eta_d N_d^+ k_B T \frac{\partial}{\partial x_i}\ln\left(\frac{N_d^+}{N_d^0 - N_d^+}\right) \tag{11}$$

The electrode is blocking for the donor current as well, which is expressed by the boundary conditions: $J_3^d\big|_{x_3=0} = 0$, $J_3^d\big|_{x_3=h} = 0$. According to Riess and Maier [71], defects in

the film with blocking electrodes are correspondent to the building defect elements due to the exhaustibility of vacancy sites limited by the parameter $N_d^0$.

In general case, kinetics of acceptors may resemble that for donors, however, in comparison with oxygen vacancies acceptor impurities are virtually immobile ($\eta_a=0$) [26], and so, $N_a^- = const$ through time and space. Typically, ferroelectric materials contain accidental metal impurities of overwhelmingly acceptor type which often take place of titanium in a perovskite unit cell [72, 73]. We will consider barium titanate unintentionally doped with acceptor impurities of molar concentrations $c_0 = 0.01$ to $0.1$ molar percent which corresponds, in SI units, to the values of $N_a^- = 1.6\times10^{24}$-$10^{25}$ [m$^{-3}$].

To deal with electromechanical properties of the ferroelectric, the description should be supported by the generalized Hooke law, where all the properties of interest are taken into account $u_{ij} = s_{ijkl}\sigma_{kl} + F_{ijkl}\dfrac{\partial P_k}{\partial x_l} + Q_{ijkl}P_k P_l$. A conventional elastic equilibrium condition

$$\frac{\partial \sigma_{ij}}{\partial x_i} = 0 \qquad (12)$$

becomes a constituent equation of the system describing the ferroelectric film. It is solved with the following boundary conditions on mechanical displacement $U_i$ and mechanical stress $\sigma_{ij}$: $(U_1 - x_1 u_m)\big|_{x_3=0} = 0$, $U_3\big|_{x_3=0} = 0$; $\sigma_{11}\big|_{x_1=h} = 0$.

The system of the above listed equations is solved by the finite-element method (FEM). The computation box captures a section of the film and the ambience, restricted by the top and bottom electrodes. The width of the box is adjusted to be equal to the period of the domain structure. Boundary conditions on the lateral sides of the computation box are periodical for each dependent variable ($P_i$, $\varphi$, $n$, $p$, $N_d^+$, $N_a^-$, $\sigma_{ij}$, $U_i$), so that their values are equal on the both sides (at $x_1=-w$ and at $x_1=w$).

# 3  Numerical results and physical analysis

## 3.1  Time evolution of stripe domain structures at different dopant concentrations

The system of Eqs. (4,6,8,10,12,16) along with the relevant boundary conditions was solved numerically for a BaTiO$_3$ (BTO) film of thickness $h = 24$ nm. Parameters used in the numerical calculations are listed in Table I. Results of the analysis are shown in Figs. 2, 3 and 4 for the cases of doping with a divalent acceptor of variable molar concentration $c_0=0.01$ mol%, $0.05$ mol% and $0.1$ mol% , respectively.

Initial donor concentrations are uniform and evaluated from electroneutrality, assuming the charge +2 of the oxygen vacancies, and corresponding to the percentage of oxygen sites being unoccupied in the following way: $N_d=1.6\times10^{24}$ m$^{-3}$ for $c_0=0.01$ mol%; $N_d=3.2\times10^{24}$ m$^{-3}$ for $c_0=0.02$ mol%; $N_d=4.8\times10^{24}$m$^{-3}$ for $c_0=0.03$ mol%; $N_d=6.4\times10^{24}$m$^{-3}$ for $c_0=0.04$ mol%; $N_d=8\times10^{24}$ m$^{-3}$ for $c_0=0.05$ mol%; $N_d=9.6\times10^{24}$ m$^{-3}$ for $c_0=0.06$ mol%; $N_d=1.2\times10^{24}$ m$^{-3}$ for $c_0=0.075$ mol%; $N_d=1.6\times10^{25}$ m$^{-3}$ for $c_0=0.1$ mol%. In the following, we also will conventionally identify spatial coordinates $x_1$ with $x$ and $x_3$ with $z$.

Fig. 2 presents time evolution of a spontaneously formed stripe domain structure for $c_0$=0.01 mol%, together with corresponding two-dimensional maps of donor concentration and bound charge density. While polarization domain pattern remains mostly unchanged (Fig. 2(a-e)), it is apparent that mobile donors substantially redistribute inside the film over time (Fig. 2(f-j)). First, they form zones underneath the top surface (Fig. 2(h)), with an excessive and depleted donor concentration, in correspondence to the electric potential distribution. It is well seen that domain walls become negatively charged along their whole length (Fig. 2(k-o)), and excessive donors are localized on them to compensate the bound charge (Fig. 2(f-j)). Formation of charged 180°-domain walls is in contrast to the previous results by Xiao et al. [33] where only 90°-domain walls got charged. This might be due to the additional account of electrostriction and flexoelectricity in our study. However, similar to [33], formation of a double charge layer is observed in the regions with polarization rotation (Fig. 2 (k-o)). As time goes on, the charged near-surface regions pull and absorb donors from neighbouring zones (Fig. 2 (i)) first and then attempt to deplete the whole film, so that the donor concentration in the bulk of the film reduces by an entire order of the magnitude (Fig. 2(j)). The redistribution in the whole film volume is explained by the fact that all the vacancies in the film, at the considered low concentration, are not able to compensate the bound charge $P_s$ and suppress the depolarization field. The acceptor concentration necessary for this can be roughly evaluated as $c^*=P_s/4eh\sim 0.1$ mol%. When the redistribution is mostly finished (Fig. 2(e)), the near-surface regions become so charged that they distort shapes of domains, albeit slightly for the considered doping of 0.01 mol%.

We note here that, due to the space charge blocking electrodes, electron subsystem is not in equilibrium with the electrode(s), and the concentration of electronic carriers remains very low everywhere. The role of electrons and holes in the migration of donors and the evolution of the domain structure is therefore virtually negligible, differently from other considerations assuming quasi-equilibrium conditions [33, 74, 75, 76].

When the doping concentration is five times increased to 0.05 mol% (Fig. 3), the changes become more evident. The redistribution of donors entails the beginning of reconstruction of the domain structure, most distinctly exhibited by recharging of the domain walls. Though not in full length, this process changes the sign of the domain wall bound charge in upper half of the walls (Fig. 3(o)). These regions become simultaneously depleted of donors more severely than the bottom half of the film (Fig. 3(j)), compare a light-yellow top half with a golden bottom half). Domain walls begin thereby to tilt so that the upper part of a domain would become slightly wedged in shape: the positive domain becomes narrower, while the negative one widens occupying more place (Fig. 3(e)).

Increasing the doping further up to 0.1 mol% (Fig. 4) we become able to observe a pronounced change of the domain structure. Depleted of donors in favour of the near-surface regions will be the whole film (Fig. 4(j)). Domain walls become tilted across the whole film thickness (Fig. 4(e)), with the bound charge of these domain walls inversed along the full length (compare Figs. 4(g) and 4(j)). The recharging process begins earlier (Fig. 4(m)) and in the end is more prominent (Fig. 4(o)), followed by the respective donor depletion (Fig. 4(j)). Furthermore, at this doping level, the significance of the account of steric effects in the course of the field-driven defect migration becomes evident. It can be observed that the peak value of the donor concentration reached near the free surface (Fig. 4(j)) achieves a magnitude close to the steric limit. The influence of steric effects on the concentration profile at different initial concentration and field values is discussed in detail in **Appendix B**.

To sum up, we see almost no changes in the domain structure at the 0.01 mol% acceptor concentration (compare Fig. 2(a) and Fig. 2(e)), slight changes at 0.05 mol% (compare Fig. 3(a), Fig. 3(d) and Fig. 3(e)), and noticeable differences at 0.1 mol% (Fig. 4(a-e)). The domain wall recharging occurs at a half film thickness at 0.05 mol% (Fig. 3(o)) and is fully completed at 0.1 mol% (Fig. 4(o)); and this process gets faster with doping increase.

As it is shown in each case, parallel stripe domain shapes change to a pattern where a negative domain becomes larger than the positive one, thus creating their wedged, or trapezoidal, shapes. It indeed speaks in favour of the negative polarization domain, pulling the average vertical polarization of the system in the negative direction as is seen in Fig. 5. This happens in competition with the variation of the space-charge accumulating regions in the near-surface layer (see Figs. 2,3,4(f-j)). The depleted region with the negative charge has always a larger height than the positively charged accumulation region, where the most of donors are collected, because of the uniform acceptor concentration. This produces an effective surface dipole layer always polarized from inside out [77], *i.e.* positively in the considered geometry. In accordance with Le Chatelier's principle, the system response strives to compensate this excess polarization of the defect origin by reconfiguring the domain structure with a prevailing negative polarization. This proceeds in a complex way, involving the varying shape of domains and of space charge zones, the recharging domain walls and also the changing donor concentration on the walls. It is worth mentioning that charge redistribution happens much faster than the rebuilding of the domain structure, which smoothens the difference between positively and negatively charged regions (compare panels (f-j) in Figs. 2-4, paying attention to the emergence and shrinking of the blue region). This makes us conclude that the polarization distribution change is a response to the defect migration and is designed to balance the charges. This mechanism is, however, much slower than the defect redistribution, which leads to a non-monotonous behaviour of characteristics like an average electric field inside the film (Fig. 6).

Thus, the time dependence of the average electric field goes through two minima and, with the increasing concentration $c_0$, even changes its asymptotic behaviour and its sign in the end for the highest concentrations. Taking the curve for the acceptor concentration $1.6 \times 10^{25}$ m$^{-3}$ ($c_0$=0.1 mol%) as an example, we can see, that emerging blue space-charge region in Fig. 4(g) at $t=2\times10^4$ s corresponds to the local maximum on the electric field curve which is followed by a minimum at $t=2\times10^5$ seconds, when the space charge begins to dilute and change its shape. At $1.5\times10^6$ seconds, the changes in the domain configuration become apparent (Fig. 4(d)), when the depleted blue region becomes narrower, but grows in its height and gives away more donors (Fig. 4(i)) thus providing recharging of adjacent domain walls (Fig. 4(n)).

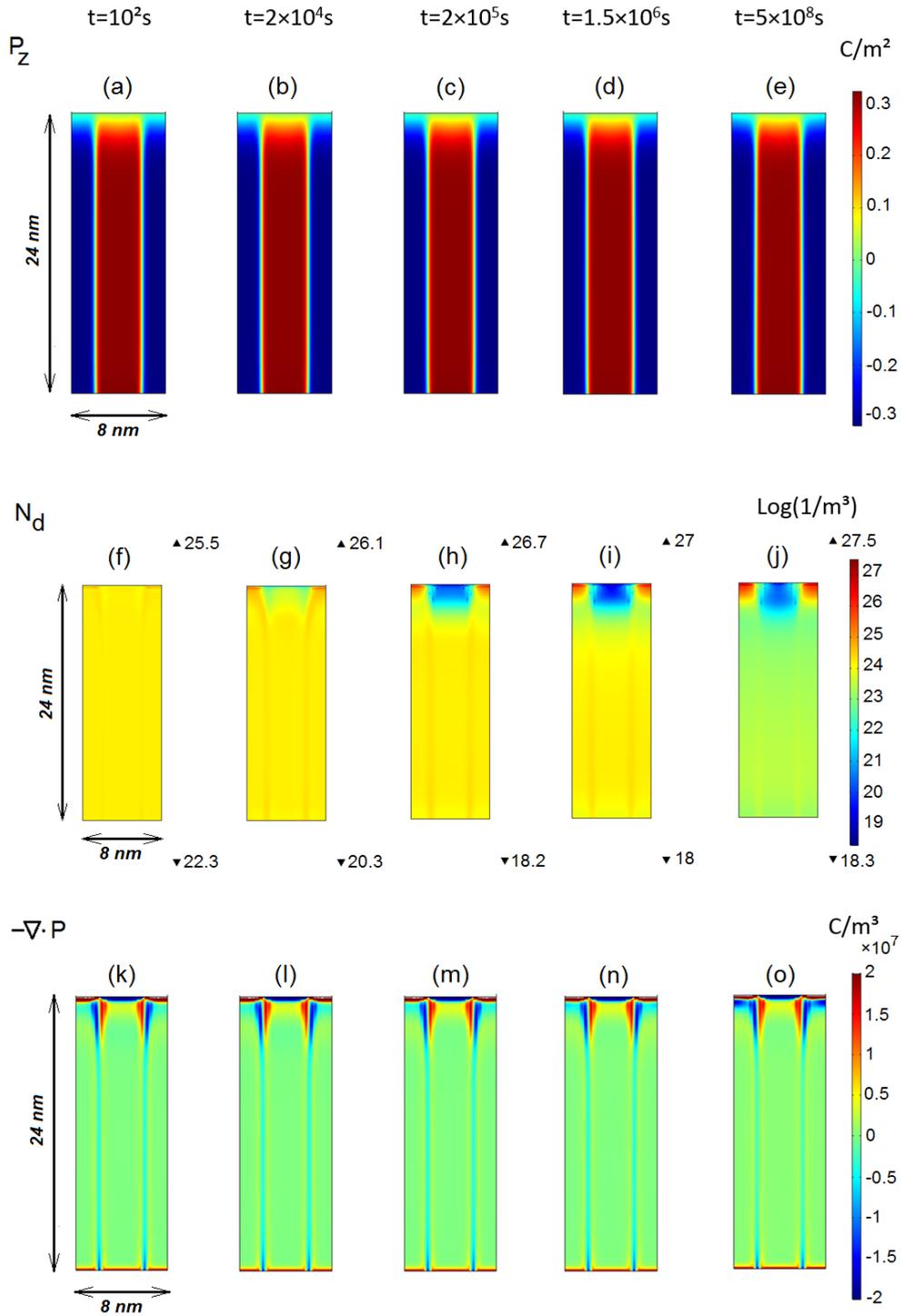

Fig. 2. Time evolution from $10^2$ s to $5\times10^8$ s (left to right) of the stripe domain structure in a 24 nm thick ferroelectric film doped with a divalent acceptor of concentration $c_0=0.01$ mol% illustrated by snapshots of two-dimensional maps of polarization $P_z$ (a-e), concentration of oxygen vacancies $N_d$ (f-j) and density of bound charge $-\nabla\cdot\mathbf{P}$ (k-o). Numbers at the top and the bottom of plots (f-j) indicate the reached limiting values of $\log(N_d)$.

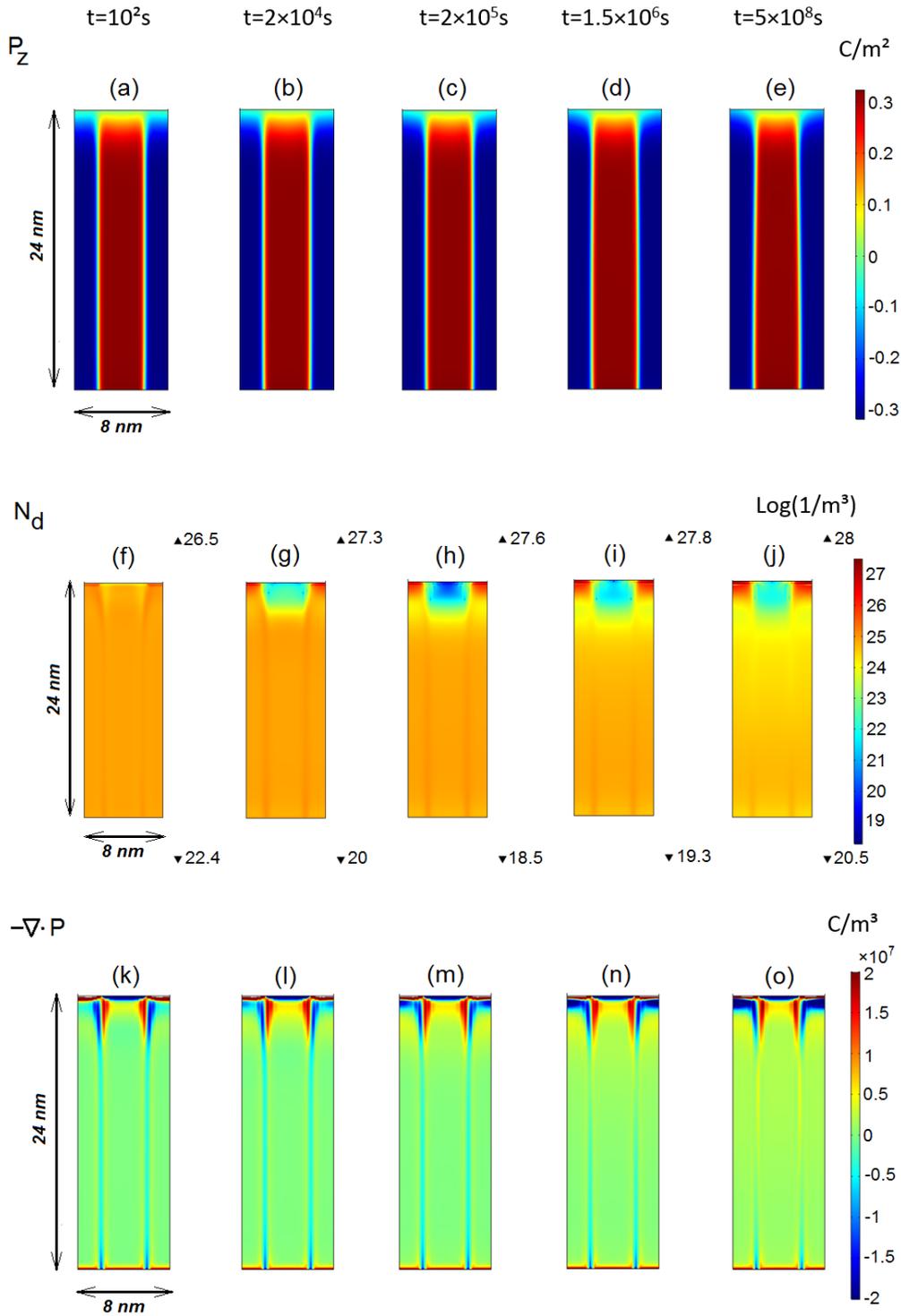

Fig. 3. Time evolution from $10^2$ s to $5\times10^8$ s (left to right) of the stripe domain structure in a 24 nm thick ferroelectric film doped with a divalent acceptor of concentration $c_0$=0.05 mol% illustrated by snapshots of two-dimensional maps of polarization $P_z$ (a-e), concentration of oxygen vacancies $N_d$ (f-j) and density of bound charge $-\nabla\cdot\mathbf{P}$ (k-o). Numbers at the top and the bottom of plots (f-j) indicate the reached limiting values of $\log(N_d)$.

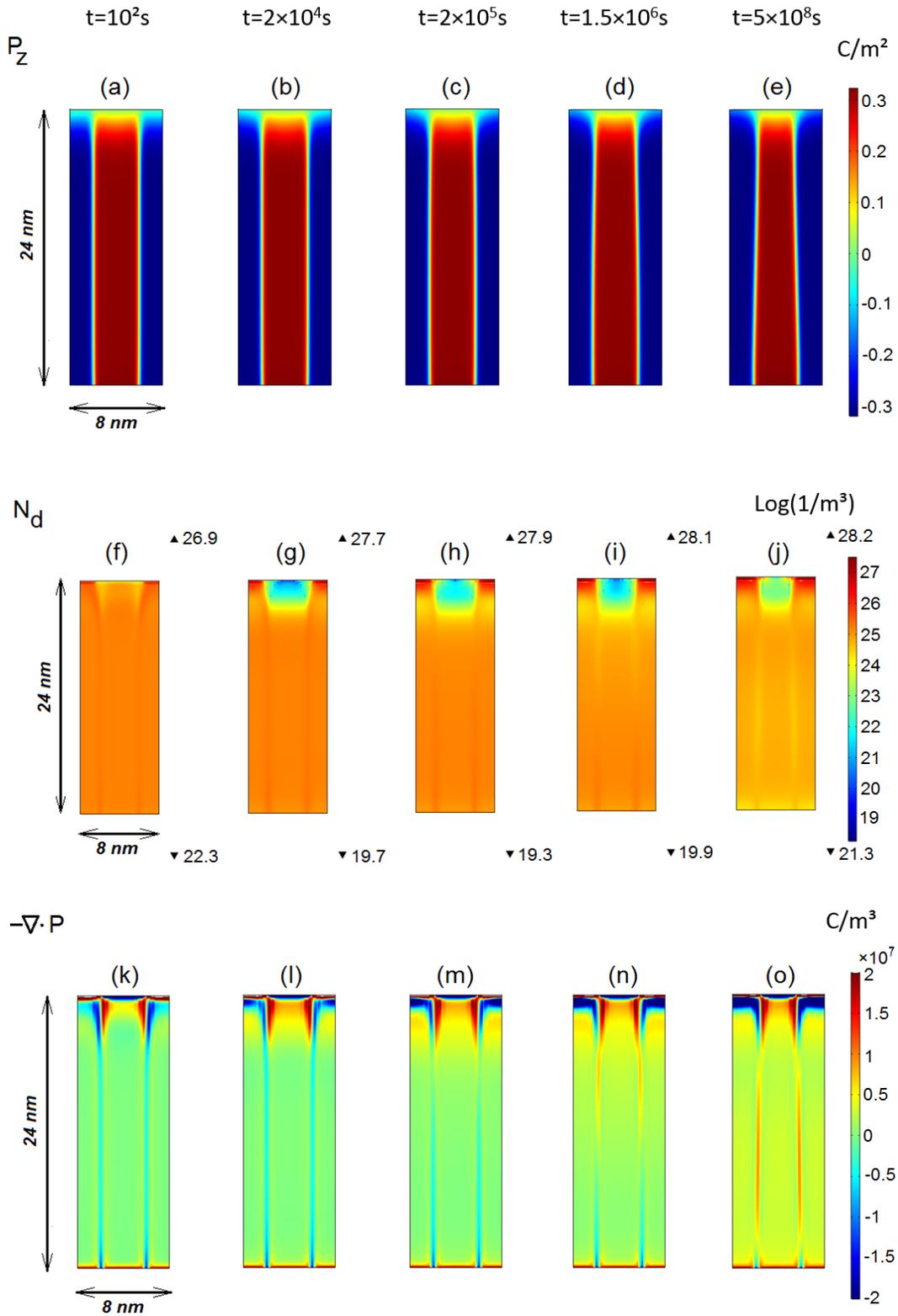

Fig. 4. Time evolution from $10^2$ s to $5\times10^8$ s (left to right) of the stripe domain structure in a 24 nm thick ferroelectric film doped with a divalent acceptor of concentration $c_0=0.1$ mol% illustrated by snapshots of two-dimensional maps of polarization $P_z$ (a-e), concentration of oxygen vacancies $N_d$ (f-j) and density of bound charge $-\nabla\cdot\mathbf{P}$ (k-o). Numbers at the top and the bottom of plots (f-j) indicate the reached limiting values of $\log(N_d)$.

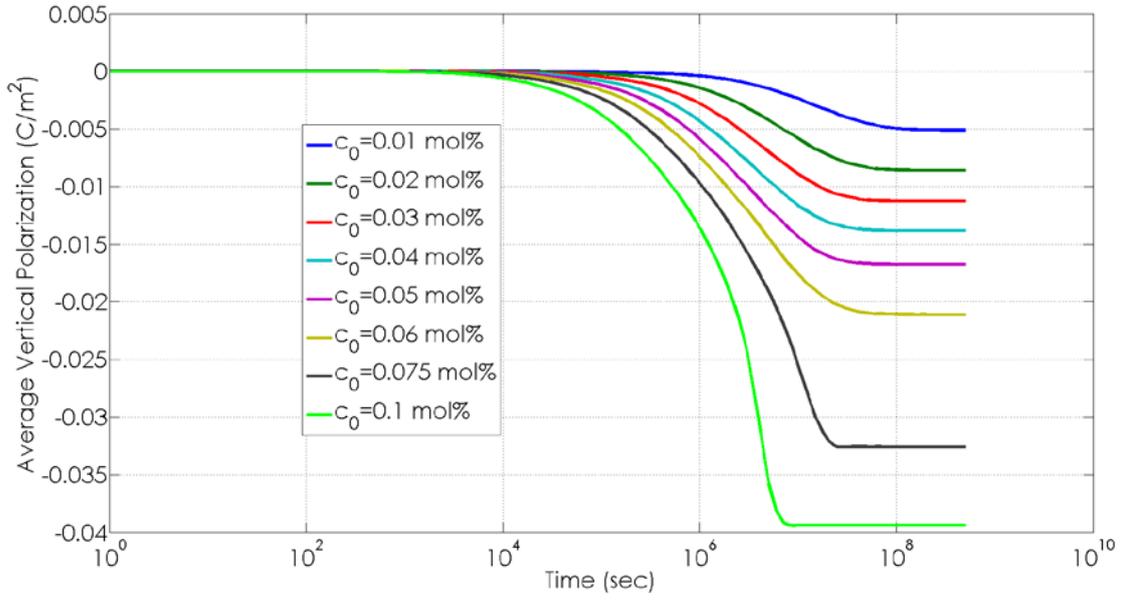

Fig. 5. Time evolution of the average vertical polarization $<P_z>$ in the 24-nm thick ferroelectric film for a set of acceptor doping concentrations $c_0$ from 0.01 mol% to 0.1 mol%.

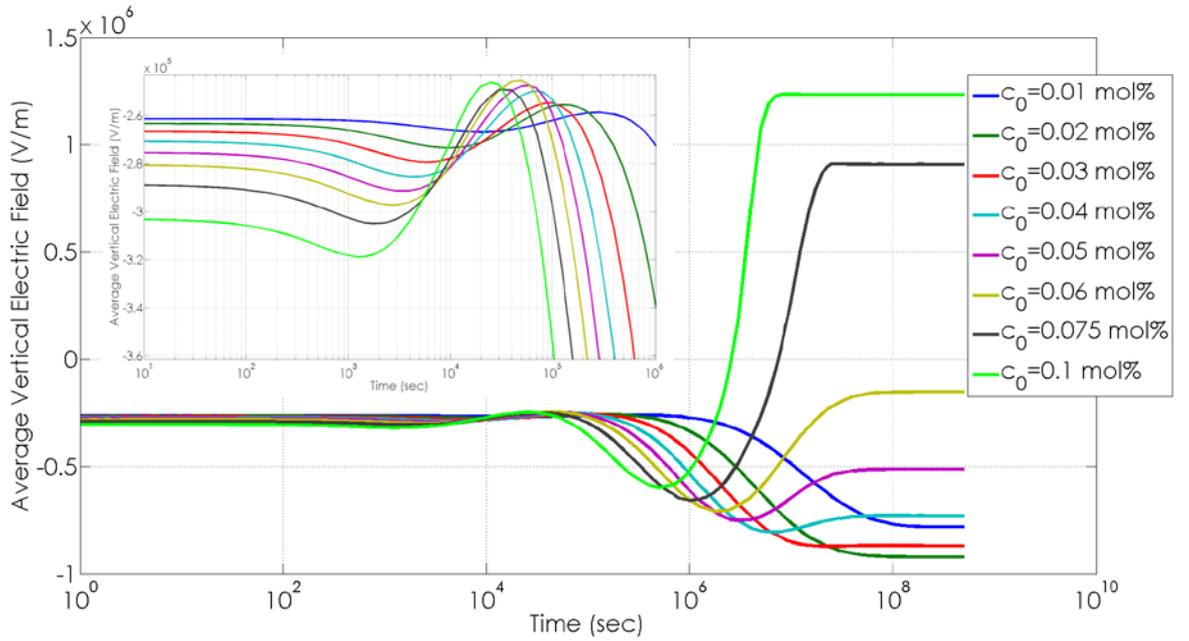

Fig. 6. Time evolution of the average electric field $<E_z>$ in a 24-nm thick ferroelectric film for a set of acceptor doping concentrations $c_0$ from 0.01 mol% to 0.1 mol%. Inset: Zoom-in of the time development from 10 to $10^6$ seconds.

At this stage the average electric field approaches zero. For doping concentrations $c_0$ of 0.03%-0.06%, the process would finish at this point (see Fig. 3(j)). For the higher concentrations, however, it continues as the time dependence of the field goes beyond zero to a stable positive level, whereby the depleted region becomes more diluted and shrinks, while donors from the bulk migrate to the red near-surface regions (Fig. 4(j)) and domain walls become almost fully recharged.

Comparing this with the time dependence of the average vertical polarization (Fig. 5), it is only seen, that the latter develops monotonically increasing in the amplitude with both the time and the concentration $c_0$ towards negative values from the very beginning with its speed increasing in correspondence to the abruptness of the average electric field changing.

We thus can talk about non-monotonous processes in the film, like donor concentration distributions (Figs. 2-4(f-j)) and average electric field evolution (Fig. 6), which are conditioned by a strife between charged defect motion and domain structure response, in contrast to monotonous processes, like average vertical polarization development (Fig. 5) conditioned by the fact that, whatever it takes, the system is heading to the most energetically favourable state. The change in the system energy, characterized by the internal bias field ($E_{ib} = \Delta G / P_s$ [12]), shown in Fig. 7 and monotonously increasing over time, is the driving force that drags the polarization structure of the system into the wedge-shaped domain form with prevalence of the down-directed polarization.

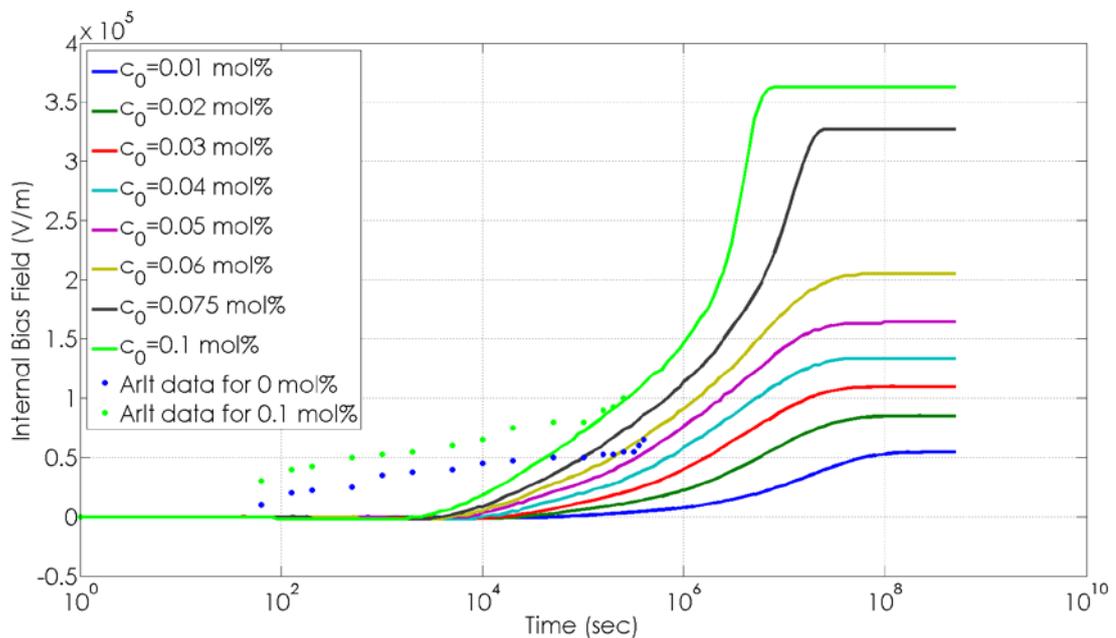

Fig. 7. Time evolution of the internal bias field $E_{ib}$ in a 24-nm thick ferroelectric film for a set of acceptor doping concentrations $c_0$ from 0.01 mol% to 0.1 mol% (solid curves) and data taken from the work [12] for Ni doping of 0 and 0.1 mol% in BT.

Internal bias field also exhibits a simple, quasi-proportional relation to the irreversible change of polarization with time, as is shown in Fig. 8. Similar observations were recently done by Fan and Tan [32] and show a fall of polarization with the bias field increase.

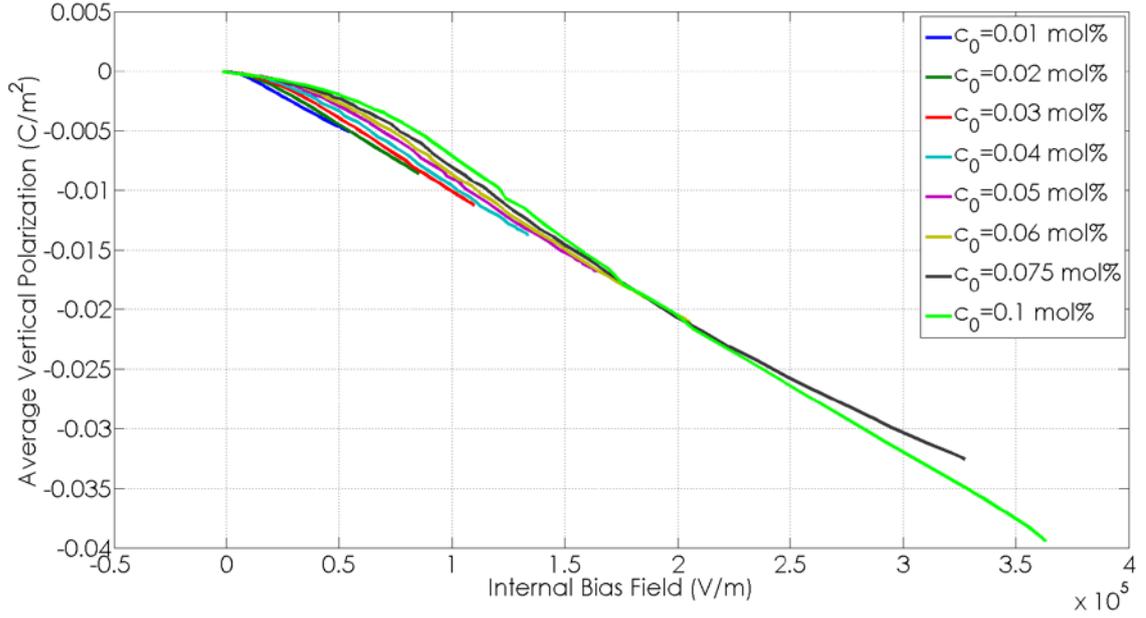

Fig. 8. Irreversible zero-field polarization dependence on internal bias field for the 24-nm ferroelectric film at acceptor doping concentrations $c_0$ from 0.01 mol% to 0.1 mol%.

## 3.2 Formation of a surface electrostatic potential: physical nature, doping and time dependences

The average electric field in the film $\langle E_z \rangle$ may be related to a surface electrostatic potential at the top surface $\varphi_s = -\langle E_z \rangle h$ being a measurable quantity. Its complicated non-monotonic behaviour with increasing time and doping concentration can be rationalized in terms of the polarization characteristics of the film. To this end, let us consider a surface polarization density (electric dipole layer) caused by the redistribution of the charged defects. By definition, the mean surface dipole moment density equals

$$p_s = \frac{1}{2w} \int_{-w}^{w} dx \int_{0}^{h} dz\, z\, \rho(x,z) \qquad (18)$$

with the free charge density given by the second term in Eq. (6a). Using the latter equation, Eq. (18) can be transformed to

$$p_s = \frac{1}{2w} \int_{-w}^{w} dx \int_{0}^{h} dz\, z \left[ \varepsilon_0 \varepsilon_b \left( \frac{\partial E_x}{\partial x} + \frac{\partial E_z}{\partial z} \right) + \left( \frac{\partial P_x}{\partial x} + \frac{\partial P_z}{\partial z} \right) \right]. \qquad (19)$$

The derivatives with respect to $x$ vanish after integration over $x$ due to periodic conditions at the sides $\pm w$. The rest terms build together a derivative of the electric displacement with respect to $z$:

$$p_s = \frac{1}{2w} \int_{-w}^{w} dx \int_{0}^{h} dz\, z\, \frac{\partial}{\partial z}\left( \varepsilon_0 \varepsilon_b E_z + P_z \right) \qquad (20)$$

Integration in parts allows one to express the polarization term in the right-hand side through the mean polarization values in the volume of the film, $\langle P_z \rangle$, and at the top surface, $\overline{P}_z(h)$.

Similarly, the electric field can be expressed through the mean value of the electric field in the film, $\langle E_z \rangle$, and the mean value at the top surface $\bar{E}_z(h)$,

$$p_s = h\left[\bar{P}_z(h) - \langle P_z \rangle + \varepsilon_0 \varepsilon_b \left(\bar{E}_z(h) - \langle E_z \rangle\right)\right], \tag{21}$$

building thus a difference between the surface and the volume mean values of the electric displacement. We note that the surface mean value of the electric displacement at the top must coincide with this value at the bottom of the film, due to the total electroneutrality, but must not vanish and may vary in time. Using the relation of the volume mean value of the electric field to the surface potential, the latter can now be expressed as

$$\varphi_s = p_s / \varepsilon_0 \varepsilon_b + h\left[\langle P_z \rangle - \bar{P}_z(h)\right] / \varepsilon_0 \varepsilon_b - h\bar{E}_z(h). \tag{22}$$

Applying Eq. (22) to the case of a fixed stripe polarization domain structure, considered in Ref. [72] in a hard-domain approximation, we note the disappearance of the both mean polarizations in the bulk and at the surface, and also the vanishing of the mean electric field at the surface for symmetry reasons in the considered geometry, thus keeping the only non-zero first term. The opposite sign of this term in Eq. (22) to the result of Ref. [77] is explained by the reference zero potential at infinity used in the latter work, differently to zero potential at the electrode adopted in the current study.

A non-monotonic behaviour of the surface potential displayed in Fig. 9 may result from the complicated interplay of different contributions in Eq. (22) having their own dynamics. Variation of the first term, representing contributions of the charged defects, is characterized by the Maxwell relaxation time of the defect migration $\tau_r = \varepsilon_0 \varepsilon_b / e Z_d \eta_d c_0$. The second term is determined by the evolution of the polarization domain structure in the volume and at the top, which seems to proceed slower. The development of the last term over time results from redistribution of the both sources of the electric field, charged defects and polarization, and thus may be characterized by another characteristic time. Fig. 9 shows a time evolution of the surface potential for a 24-nm film with acceptor doping concentrations varying from 0.01 mol% to 0.1 mol%. The main properties of the curves and their immediate correspondence to charge distributions over the film bulk were described in the section 3.1. Here, we will analyse these dependencies in an attempt to devise mechanisms governing their behaviour.

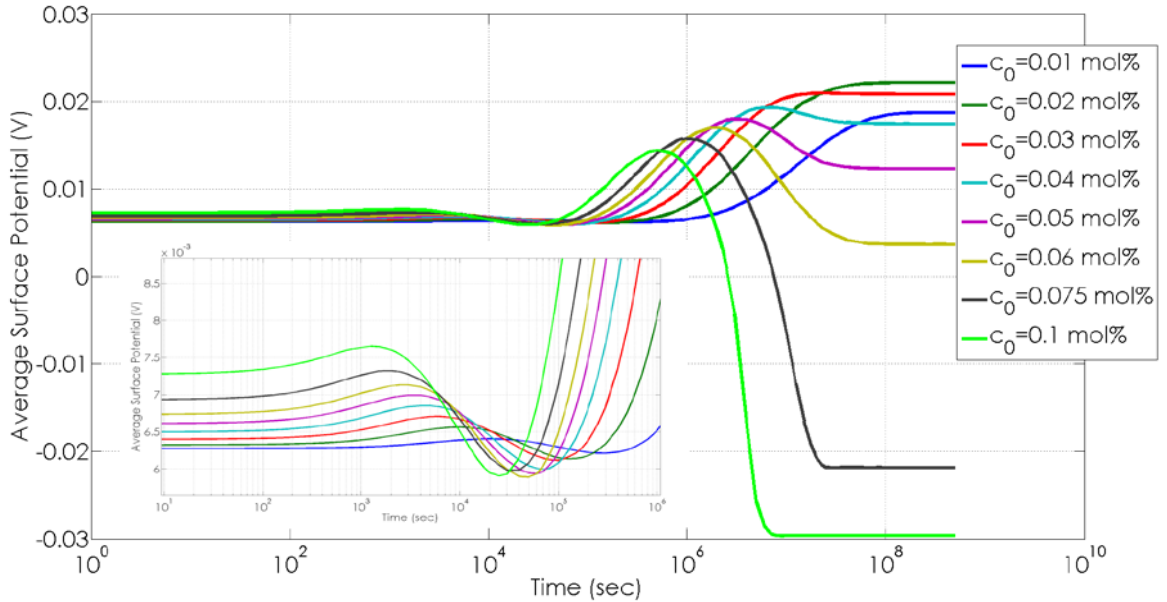

Fig. 9. Time evolution of the average surface potential in a 24-nm thick ferroelectric film for a set of acceptor doping concentrations $c_0$ from 0.01 mol% to 0.1 mol%. Inset: Zoom-in of the time development from 10 to $10^6$ seconds.

In Fig. 10, inflection points found on the curves are designated using the time dependence of the surface potential for doping of 0.1 mol% as an example. Thus, four characteristic times are found representing different specific tendencies in curve's behaviour. Leading to the first small local maximum is $\tau_0$, followed by $\tau_1$, when the dependency is driven to a small local minimum. This escalates, when the curve rises to its peak with an inflexion point seen at $\tau_2$ and, finally, culminates when the average surface potential falls to zero and beyond with the characteristic time of $\tau_3$ until it approaches a constant negative value.

This behaviour strongly hints at the existence of at least two mechanisms competing with each other with one of them prevailing in a long run. From the above-mentioned processes, those two mechanisms can be identified as the donor redistribution inside the film, creating nonuniform space-charge areas, and polarization response which smooths and compensates charge inhomogeneities by adjusting the domain structure. One way to prove this understanding is to present a concentration dependence of the characteristic times and compare it with the Maxwell time for donors. These comparative dependencies are shown in Fig. 11 where the similarity between the Maxwell time and $\tau_2$ concentration dependencies is clearly visible. From this and previous analyses it can be stated that the elevation of the average surface potential to its maximum is linked to the defect redistribution whereas the final fall of the curve results from the domain structure response.

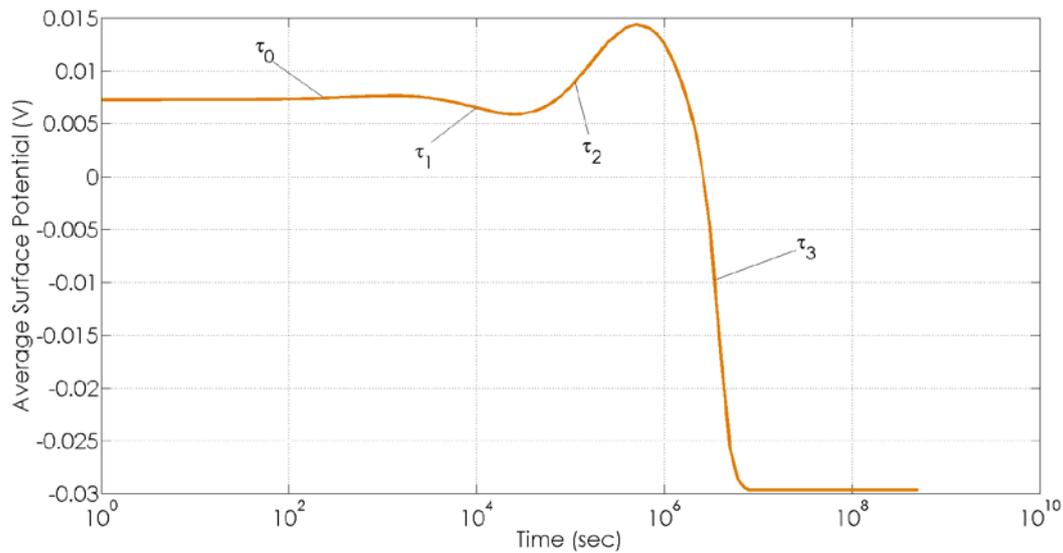

Fig. 10. Characteristic times marked on the average surface potential time-evolution curve from Fig. 9 for acceptor doping concentration $c_0$ of 0.1 mol%.

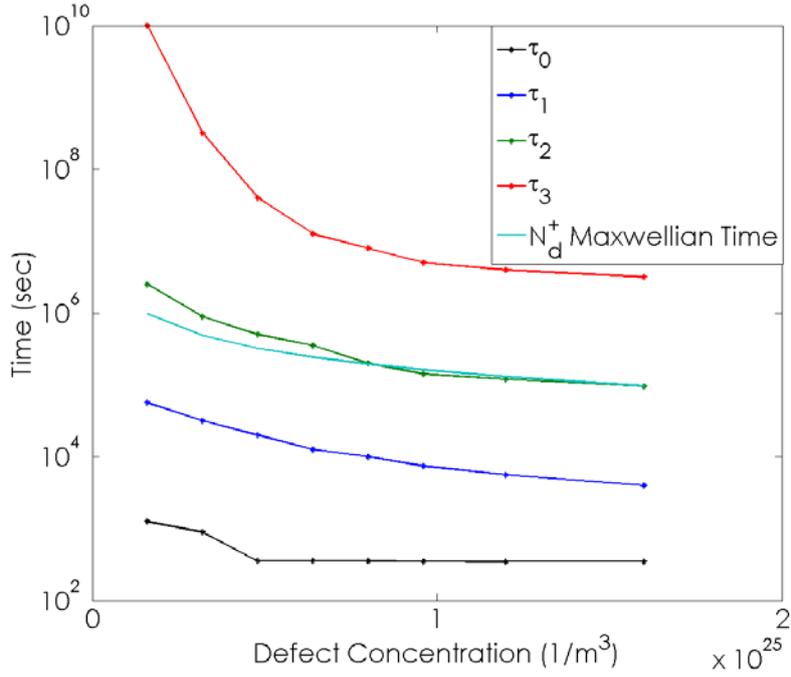

Fig. 11. Characteristic times doping concentration dependencies for each of the curves from Fig. 9, compared with the acceptor doping concentration dependence of the Maxwell relaxation time for donors.

A way to make sure that the mechanisms have been identified correctly is to switch off one of them and investigate the average surface potential behaviour influenced by the remaining one. For that we turn to the case similar to the previously studied fixed domain structure when only dopant distribution and electronic densities are subjects to change [77]. Figure 12 shows a time dependence of the average surface potential when the domain structure, self-organized in our geometry with the accounted physical mechanisms, does not change anymore. Indeed, when the polarization response is eliminated, the influence of the redistributed charge remains undisputed and finishes only by its saturation. The final potential drop with its characteristic time $\tau_3$ exists no more, and $\tau_2$, linked to the Maxwell time for donors, is all that makes noticeable changes to the surface potential state.

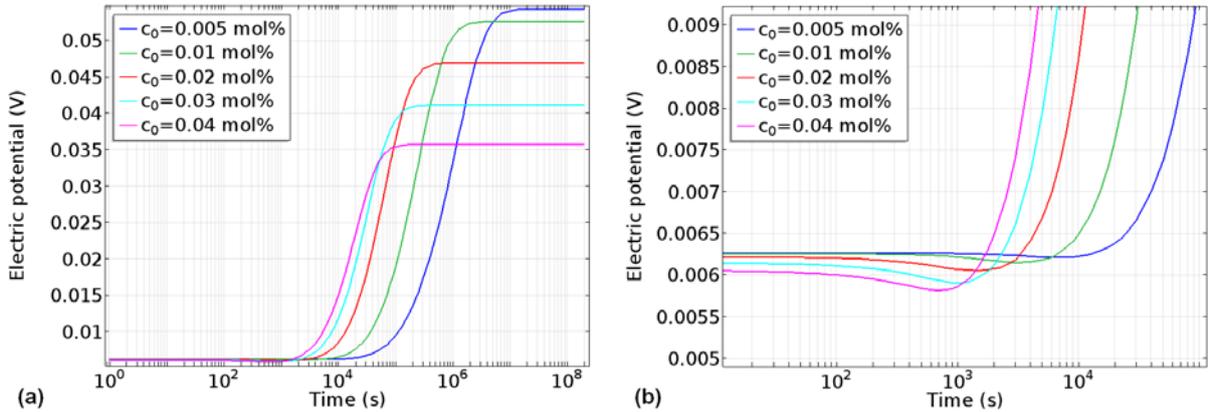

Fig. 12. (a): Time evolution of the average surface potential in a 24-nm thick ferroelectric film in the case without rearrangement of the domain structure for the acceptor doping concentrations $c_0$ from 0.005 mol% to 0.04 mol%. (b): Zoom-in of the time-development from 10 to $10^5$ seconds.

When we, however, zoom in to see what happens at the beginning of the average surface potential development, we observe a very small local minimum, formed by a process characterized by the time $\tau_1$ that drives the curve downwards. As nothing else of peculiarities is found on the curve, in the end, when no domain structure change is allowed, there are only two processes left: one characterized by $\tau_1$ and another characterized by $\tau_2$. These times might be related to a very nonuniform electric field distribution within the domain structure, with migration processes of different velocities in the near-surface area and deep in the film.

Previously, the characteristic times were estimated roughly by observing the inflexion points for each tendency occurring on the average surface potential dependence on time. Due to competitive processes, positions of inflexion points might be thus shifted due to distortion of the segments which could influence each other. To recreate parameters of each segment we can fit the model dependencies to curves constructed analytically with a set of parameters mimicking the behaviour of the average surface potential. The parametrized expression for such dependencies denoted as $V(t)$ reads as follows

$$V(t) = m_0 + (m_1 - m_0)\left(1 - \exp\left(\frac{-t}{\tau_0}\right)\right) + (m_2 - m_1)\left(1 - \exp\left(\frac{-t}{\tau_1}\right)\right) + (m_3 - m_2)\left(1 - \exp\left(\frac{-t}{\tau_2}\right)\right) +$$
$$+ (m_4 - m_3)\left(1 - \exp\left(\frac{-t}{\tau_3}\right)\right)$$

(23)

where $m_0$, $m_1$, $m_2$, $m_3$, $m_4$ are asymptotic levels to which the correspondent sections tend, and $\tau$ are the characteristic times of the processes. Fitting curves for the average surface potential dependencies at three doping concentrations $c_0$ = 0.01, 0.04, 0.05, 0.06 and 0.1 mol% are shown in Figure 13. Each fitting $V(t)$ plot contains its own set of parameters mi and $\tau_i$. Being extracted from the expression, all characteristic times are shown in Figure 14 in dependence of dopant concentration. In this case, the Maxwell time for donors does not coincide with a certain characteristic time but lies right in between $\tau_1$ and $\tau_2$. In fact, on the logarithmic time scale, a geometric mean between $\tau_1$ and $\tau_2$ fits the Maxwell time for donors pretty well, indicating that the coupled processes 1 and 2 are both driven by the donor migration.

By analysing the behaviour of the average surface potential curve, when the domain structure response is allowed (Fig. 9) or prohibited (Fig. 12) and observing concentration dependencies of its characteristic times (Fig. 14), we may identify the mechanisms behind its non-monotonous time dependence. Namely, within a concept of the mechanism competition, the effect of dopant redistribution is responsible for processes characterized by times $\tau_1$ and $\tau_2$, ultimately pulling the average surface potential to its maximum value, whereas the motion of domains can be characterized by the times $\tau_0$ and $\tau_3$ (which both vanish should we prohibit the domain reconstruction) leading ultimately to the drop of the average surface potential. With the doping concentration increase, all characteristic times gradually decrease, and the effect of the domain structure response becomes ever stronger, pulling the potential down to zero and beyond.

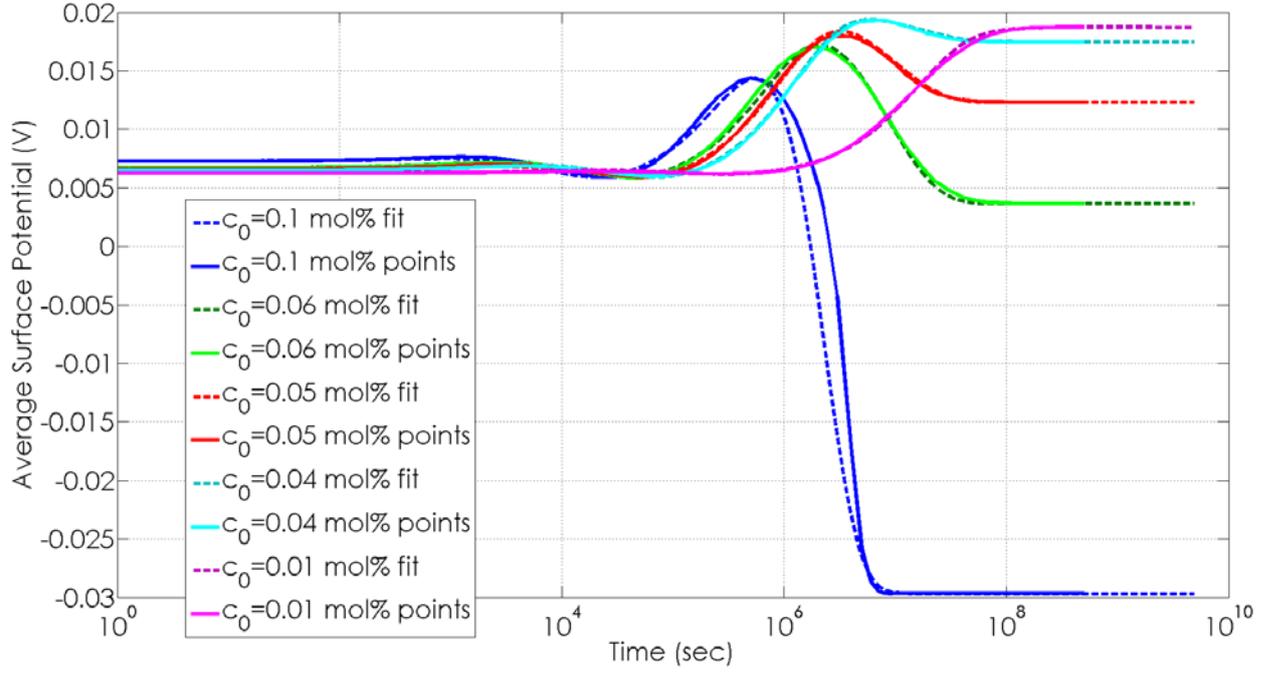

Fig. 13. Fitting of average surface potential time-dependence curves for acceptor doping concentrations of 0.01 (purple), 0.04 (cyan) 0.05 (red), 0.06 (green) and 0.1 (blue) mol % in a 24-nm ferroelectric film. Dependencies obtained from the modelled points are shown with solid lines, fitting curves $V(t)$ are shown with dashed lines.

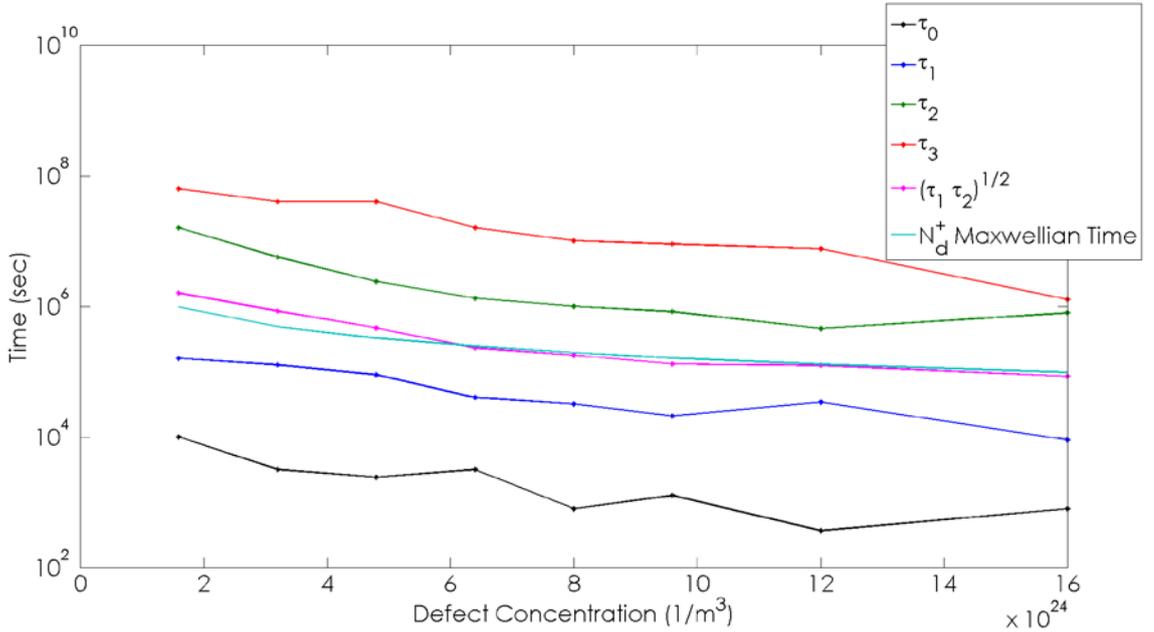

Fig. 14. Acceptor doping concentration dependencies of characteristic times identified as parameters of $V(t)$ fitting curves shown in Fig. 13, compared to the Maxwell time for donors and geometric mean between $\tau_1$ and $\tau_2$.

The asymptotic value of the surface potential displays a non-monotonic dependence on the doping concentration that may be observed in Fig. 9. In the system with the variable domain structure, it exhibits a maximum at $c_0 \cong 0.02$ mol% and in the system with a fixed domain structure at $c_0 \cong 0.005$ mol% (Figure 15). For the latter case, the concentration dependence of the surface potential, proportional to the mean surface dipole moment density, can be roughly estimated from a simple reasoning using the hard domain approximation [77]. For very low acceptor concentrations, $c_0 \ll c^*$, the whole film area becomes eventually depleted of oxygen vacancies by at least one order of the magnitude, as is seen in Fig. 2(j). Assuming that the resulting positive space charge of $qc_0 2wh$ is mostly concentrated close to the negative domain face, this produces a mean surface dipole moment density of $p_s \approx eh^2 c_0$, increasing with $c_0$. For high acceptor concentrations, $c_0 \gg c^*$, the area of depth $s \approx P_s/2qc_0$ provides enough vacancies to fully screen the negative surface charge density $P_s$. This produces a mean surface dipole moment density of $p_s \approx P_s^2/16ec_0$, decreasing with $c_0$. The maximum of this dependence is located about the doping concentration $c_{max} \cong P_s/4eh$ which coincides with $c^*$ of about 0.1 mol%. This value overestimates the numerical results by about one order of the magnitude because it oversimplifies the spatial charge distribution and neglects the spatial polarization distribution in the fixed domain structure.

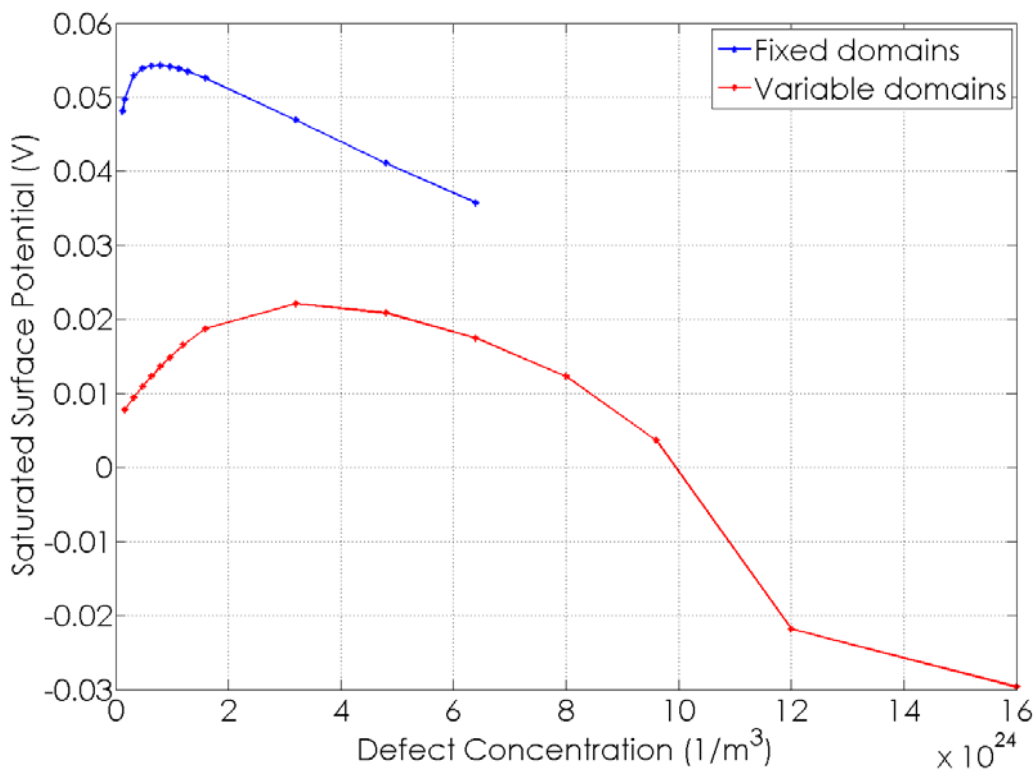

Fig. 15. Acceptor concentration dependence of the saturated average surface potential for 24-nm ferroelectric film with fixed (blue curve) and variable (red curve) domain structures.

## 4 Discussion and Conclusions

In this work, we have self-consistently studied the coupled evolution of the polarization domain structure and the distribution of the mobile charge defects in the acceptor-doped barium titanate, representative of a wide class of perovskite ferroelectrics. The phase-field simulations based on the time-dependent LGD approach revealed a hierarchy of characteristic times related to different coupled processes in the considered complicated system.

The effect of nano-sized static defects and charge distributions on the domain formation and development in undoped tetragonal lead titanate under the changing electric field was previously studied by phase-field simulations in Refs. [78, 79]. In other phase-field simulations of undoped tetragonal lead titanate [80] it was shown that introduction of surface charges may produce bimodal domain structures as well as wedge-shaped domains, similar to those observed by TEM in a broad compositional range of PZT [81]. Phase-field simulations of tetragonal barium titanate doped with oxygen vacancies demonstrated formation of charged 90°-domain walls leaving 180°-domain walls uncharged [33]. In contrast to the above-mentioned simulations, in this work we considered self-consistently a time-evolution of a self-organized domain structure coupled with the migration of mobile oxygen vacancies in an acceptor-doped tetragonal barium titanate without introduction of artificial excess charges. Among other effects, the formation of charged 180°-domain walls was established which might be due to the account of the electrostriction and flexoelectricity in contrast to the previous studies [33, 78, 79, 80]. The development of the system may be divided in the following subsequent stages:

- Primary process: depolarization field-driven migration of oxygen vacancies leading to the formation of asymmetric alternating space charge zones at the top of the stripe domains which are characterized by an effective surface dipole layer;
- Secondary process: a domain structure reconstruction by tilting the domain walls and breaking the vertical polarization balance to oppose the surface dipole layer;
- Tertiary effects: re-charging of the tilted domain walls and the following accumulation of charge defects therein.

Formation of the trapezoidal domains in the course of the domain re-construction reminds of the wedge-shaped domains observed by TEM and simulated in Ref. [80]. Furthermore, in recent *in-situ* TEM studies of a BZT–BCT ceramic, the macroscopic aging features, such as the development of the internal bias field and the degradation in switchable polarization were for the first time correlated with the visualized microscopic domain wall clamping and domain disruption resulted from the accumulation of oxygen vacancies driven by depolarization field [32]. Some experimentally observed features may be related to those revealed in simulations of our model domain structure. Thus, a linear reduction of the switchable polarization with the increasing internal bias field in the experiment [32] may be related to the build-up of the zero-field negative polarization in our modeled stripe domain structure also exhibiting such a linear dependence (Fig. 8). Moreover, the linear experimental dependence of the arising internal bias field for intermediate times in a semi-logarithmic presentation [32] is in agreement with the simulated behavior in Fig. 7.

Characteristic amplitude of the arising internal bias field is in a reasonable quantitative agreement with the available aging measurements on the acceptor doped barium titanate [12], as is seen in Fig. 7. It is interesting to observe that the nominally undoped sample also exhibits noticeable internal bias field evolving with time. This is due to the fact that the referred acceptor concentration accounts only the intentional Ni doping leaving aside the omnipresent accidental metal impurities [73, 74]. Comparison with simulating curves suggests a presence of metal impurities with concentration of about 0.06 mol%. The time dependence of the internal bias field in the experiment is, however, by three orders of the magnitude faster than that in simulations. This may be explained by the presence of defect dipoles, not considered in our simulations, which contribute to the volume aging effect and possess substantially shorter characteristic aging times than those of the long-range vacancy migration [14, 22, 23].

# Acknowledgements

This work was supported by the Deutsche Forschungsgemeinschaft (DFG) via the grant No. 405631895 (GE-1171/8-1).

# Appendix A. Model Parameters

Table AI. Description of coefficients, tensors, constants and parameters used in the calculations

| Description | Designation | Value | Units |
|---|---|---|---|
| Main parameters | | | |
| Film thickness | $h$ | 24 | nm |
| Temperature (for all temperature-dependent parameters) | $T$ | 300 | K |
| Background dielectric permittivity | $\varepsilon_b$ | 7 | dimensionless |
| Ambience permittivity | $\varepsilon_e$ | 1 | dimensionless |
| Vertical permittivity | $\varepsilon_{33}$ | 56 | dimensionless |
| Lateral permittivity | $\varepsilon_{11}$ | 2200 | dimensionless |
| Misfit strain | $u^{Sm}$ | –0.01 | dimensionless |
| Surface screening length | $\lambda$ | $\gg 1$ | nm |
| Landau expansion coefficient | $\alpha_{ij}$ | $\alpha_1 = -2.71 \times 10^7$ | $\dfrac{m}{F}$ |
| Landau expansion coefficient | $\beta_{ijkl}$ | $\beta_{11} = -6.38 \times 10^8$<br>$\beta_{12} = 3.23 \times 10^8$ | $\dfrac{m^5}{FC^2}$ |
| Landau expansion coefficient | $\gamma_{ijklmn}$ | $\gamma_{111} = 7.89 \times 10^9$<br>$\gamma_{112} = 4.47 \times 10^9$<br>$\gamma_{123} = 4.91 \times 10^9$ | $\dfrac{m^9}{FC^4}$ |
| Gradient coefficients | $g_{ijkl}$ | $g_{11} = 5.1 \times 10^{-10}$<br>$g_{12} = -0.2 \times 10^{-10}$<br>$g_{44} = 0.2 \times 10^{-10}$ | $\dfrac{m^3}{F}$ |
| Khalatnikov coefficient | $\Gamma$ | $10^5$ | $\dfrac{s\,m}{F}$ |
| Electrostriction tensor | $Q_{ijkl}$ | $Q_{11} = 0.11$<br>$Q_{12} = -0.045$<br>$Q_{44} = 0.029$ | $\dfrac{m^4}{C^2}$ |
| Mechanical compliance tensor | $s_{ijkl}$ | $s_{11} = 8.3 \times 10^{-12}$<br>$s_{12} = -2.7 \times 10^{-12}$<br>$s_{44} = 9.24 \times 10^{-12}$ | $\dfrac{1}{Pa}$ |
| Donor and acceptor ionization | $Z_d$ and $Z_a$ | 2 | dimensionless |

| | | | | |
|---|---|---|---|---|
| degrees | | | | |
| Flexoelectric tensor | $F_{ijkl}$ | $F_{11} = 2.46 \times 10^{-11}$ $F_{12} = 0.48 \times 10^{-11}$ $F_{44} = 0.05 \times 10^{-11}$ | $\dfrac{m^3}{C}$ | |
| Effective masses of electron and hole | $m_n, m_p$ | $m_n = 1.2 m_e$ $m_p = 8 m_e$ | $kg$ | |
| Densities of states in the conduction and valence bands | $N_C, N_V$ | $N_C = 1.65 \times 10^{25}$ $N_V = 2.84 \times 10^{26}$ | $m^{-3}$ | |
| Steric limit of defect concentration | $N_d^0, N_a^0$ | $2 \times 10^{28}$ | $m^{-3}$ | |
| Electron mobility | $\eta_e$ | $6.9 \times 10^{-11}$ | $\dfrac{m^2}{V\,s}$ | |
| Hole mobility | $\eta_h$ | $4.25 \times 10^{-11}$ | $\dfrac{m^2}{V\,s}$ | |
| Donor defect mobility | $\eta_d$ | $10^{-21}$ | $\dfrac{m^2}{V\,s}$ | |
| Acceptor defect mobility | $\eta_a$ | $0$ | $\dfrac{m^2}{V\,s}$ | |
| Energy levels (Vacuum level used as zero point) | | | | |
| Bottom of the conduction band | $E_C$ | -3.8 | eV | |
| Top of the valence band | $E_V$ | -7 | eV | |
| Donor level | $E_d$ | -4.05 | eV | |
| Acceptor level | $E_a$ | -6.75 | eV | |
| Bandgap width | $E_g$ | 3.2 | eV | |
| Constants | | | | |
| Boltzmann constant | $k_B$ | $1.38 \times 10^{-23}$ | $\dfrac{J}{K}$ | |
| Elementary charge | $e$ | $1.6 \times 10^{-19}$ | $C$ | |
| Electron mass | $m_e$ | $9.11 \times 10^{-31}$ | $kg$ | |
| Planck constant | $\hbar$ | $1.055 \times 10^{-34}$ | $J \cdot s$ | |
| Characteristic times | | | | |
| Characteristic response time of the domain structure | $\dfrac{\Gamma}{\alpha_{ij}}$ | $3.69 \times 10^{-3}$ | $s$ | |
| Maxwellian time for donors | $\dfrac{\varepsilon_0 \varepsilon_{33}}{2 e \eta_d N_d^+}$ | $10^5 \div 10^6$ | $s$ | |
| Maxwellian time for electrons | $\dfrac{\varepsilon_0 \varepsilon_{33}}{2 e \eta_e n}$ | $2.2456 \times 10^{14}$ | $s$ | |
| Maxwellian time for holes | $\dfrac{\varepsilon_0 \varepsilon_{33}}{2 e \eta_h p}$ | $3.6491 \times 10^{14}$ | $s$ | |

# Appendix B. Steric effect on the defect distribution

To expose the steric effect on the concentration profile of oxygen vacancies in the course of their field-driven accumulation near the free sample surface we consider here a one-dimensional drift-diffusion problem assuming for simplicity a driving force induced by a constant positive electric field $E$. This simplifies the current density expression (13) to

$$J^d = eZ_d \eta_d N_d^+ E - \eta_d N_d^+ k_B T \frac{d}{dz} \ln\left(\frac{N_d^+}{N_d^0 - N_d^+}\right) \quad (A.1)$$

Using this form, the continuity equation (12) for the evolving concentration $N_d^+(z,t)$ will be solved numerically considering different initial uniform concentrations of donors $N_d^I$ and different values of the applied constant field $E$. As boundary conditions in a final computation domain $0 < z < h$ we use the blocking conditions (14) specialized to the one-dimensional case, $J^d(0,t) = J^d(h,t) = 0$. These conditions provide conservation of the total number of defects, given by the initial condition,

$$\int_0^h dz N_d^+(z,t) = N_d^I h . \quad (A.2)$$

Fortunately, the final equilibrium distribution of the donors resulting from the concentration profile evolution can be obtained in a closed form from the requirement $J^d = 0$. This will be done first to be able to justify later the numerical solutions. The equilibrium condition following from Eq. (A.1) reads

$$\frac{d}{dz} \ln\left(\frac{N_d^+}{N_d^0 - N_d^+}\right) = \frac{R}{h} . \quad (A.3)$$

with a dimensionless constant $R = eZ_d E h / k_B T$. This equation can be directly integrated resulting in a solution

$$N_d^+ = \frac{N_d^0}{1 + A e^{-Rz/h}} \quad (A.4)$$

with an arbitrary constant $A$. The latter constant can be found from the normalization condition (A.2) to equal

$$A = \frac{e^R - e^{R\nu}}{e^{R\nu} - 1} \quad (A.5)$$

with the initial fraction of oxygen vacancies $\nu = N_d^I / N_d^0$. Examples of equilibrium distributions (A.4) for different initial uniform concentrations and different values of the applied field are displaced in Fig. A1.

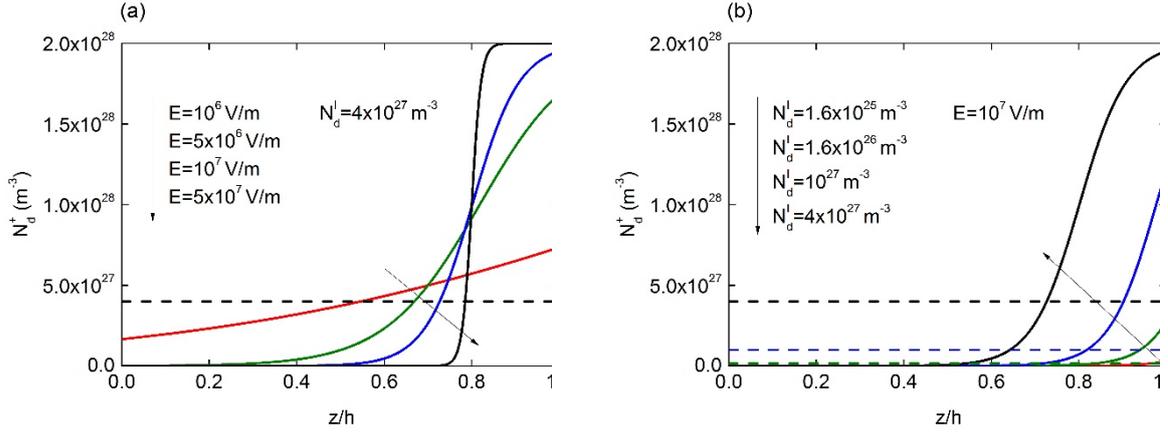

Fig. S1. Equilibrium profiles of oxygen vacancy concentration at the end of the redistribution with account of steric effects (a) for a fixed magnitude of the initial uniform concentration and different values of the applied constant electric field as indicated in the plot and (b) for a fixed value of the applied electric field and different magnitudes of the initial uniform concentration as indicated in the plot.

The evolution of the concentration profile with time can be obtained by solving Eq. (12) using the expression for the current density (A.1). Numerical solutions of this equation are exemplarily shown in Fig. A2 for high (Fig. A2 (a)) and low (Fig. A2 (b)) initial concentrations.

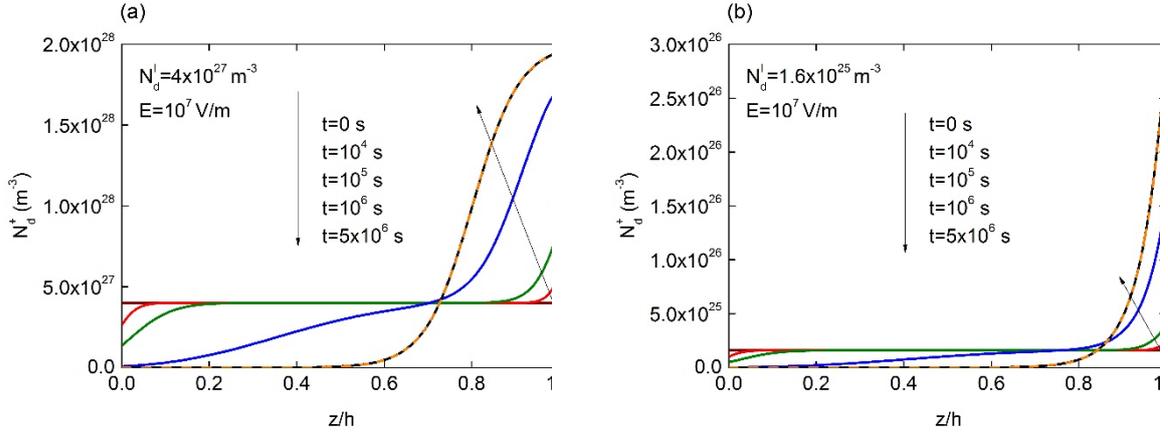

Fig. S2. Time-dependent profiles of the oxygen vacancy concentration in the course of the field-driven redistribution for the high (a) and low (b) initial uniform concentrations for the same applied field as indicated in the plots. Analytical solutions for the corresponding equilibrium states, shown by dashed lines, coincide identically with the numerical solutions at the latest stage.

It is useful to compare the concentration profile (A.4) with that resulting from the drift-diffusion equation without account of the steric effects. In this case the equilibrium condition changes to

$$J^d = eZ_d \eta_d N_d^+ E - \eta_d N_d^+ k_B T \frac{\partial N_d^+}{\partial z} = 0 \qquad (A.6)$$

with an obvious solution

$$N_d^+ = N_d^I \frac{R e^{Rz/h}}{e^R - 1} \qquad (A.7)$$

normalized by the same condition (A.2). The solutions (A.4) and (A.7) can be compared considering Figs. A1 and A3 using the same values of the initial donor concentration and different electric field values.

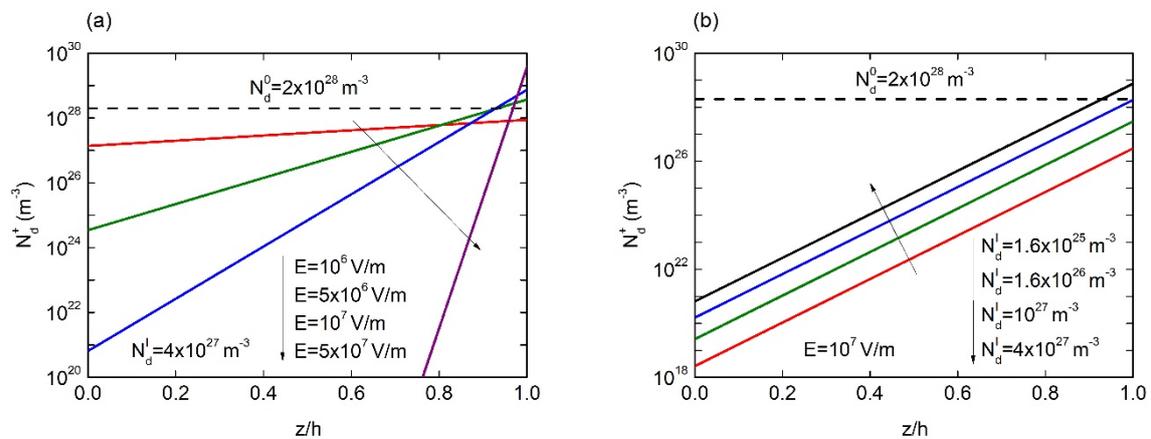

Fig. S3. Equilibrium profiles of oxygen vacancy concentration at the end of the redistribution without steric effects (a) for a fixed magnitude of the initial uniform concentration and different values of the applied constant electric field as indicated in the plot and (b) for a fixed value of the applied electric field and different magnitudes of the initial uniform concentration as indicated in the plot.

Saturation of the concentration due to steric effects near the surface $z=h$ at the peak value $N_d^0$ is apparent at high applied fields (Fig. A1(a)) as well as at a high initial concentration (Fig. A1(b)). Without steric limitations the concentration in the peak region exceeds the value $N_d^0$ at the field values relevant to the problem (Fig. A3(a)) for the same initial concentration as in Fig. A1(a). The steric limit is exceeded also for the initial concentration by one order of the magnitude smaller than $N_d^0$ (Fig. A3(b). It can be observed that, for the highest acceptor concentration of 0.1 mol% considered in this work, the peak value of the donor concentration reached near the free surface (Fig. 4(j)) achieves a magnitude close to the steric limit.